\documentclass[runningheads]{llncs}
\usepackage{graphicx}
\begin{document}
\title{Twitter Interaction to Analyze \\ 
Covid-19 Impact in Ghana, Africa \\
from March to July}
\titlerunning{Twitter Analysis, Covid-19, Ghana, March-July}
\author{Josimar E. Chire Saire\inst{1} \and Kobby Panford-Quainoo\inst{2}}
\authorrunning{Chire Saire, J. and Panford-Quainoo K. }
%
\institute{Institute of Mathematics and Computer Science (ICMC),\\  University of São Paulo (USP), Sao Carlos, SP, Brazil\\ 
\and {African Masters in Machine Intelligence (AMMI) - \\African Institute for Mathematical Sciences (AIMS) \\
Kigali, Rwanda} \\
\email{jecs89@usp.br, kpanford-quainoo@aimsammi.org}
}
\maketitle              
\begin{abstract}


The novel coronavirus, COVID-19, has impacted various aspects of the world from tourism, business, education, and many more. Like for every country, the global pandemic has imposed similar effects on Ghana. During this period, citizens of this country have used social networks as a platform to find and disseminate information about the infectious disease and also share their own opinions and sentiments. In this study, we use text mining to draw insights from data collected from the social network, Twitter. Our exploration of the data led us to understand the most frequent topics raised in the Greater Accra region of Ghana from March to July 2020. 
We observe that the engagement of users of this social network was initially high in March but declined from April to July. The reason was probably that the people were becoming more adapted to the situation after an initial shock when the disease was announced in the country. We also found certain words in these users’ tweets that enabled us to understand the sentiments and mental state of individuals at the time.

\end{abstract}

\section{Introduction}
The global pandemic, COVID-19, formally known as the severe acute respiratory syndrome coronavirus 2 (SARS-COV-2) \cite{Zheng2020}, was declared a pandemic and a global threat by the World Health Organization (WHO) on 11-03-2020 \cite{whowebpage}. 

Since December 2019, when its outbreak was first announced in the city of Wuhan in China, the disease has spread very rapidly to many other countries. As of 11-03-2020, when the disease was declared a global pandemic, a total of 6,546 cases and 355 deaths had been recorded by WHO \cite{whodashboard}.

\begin{figure}[hbpt]
\centerline{
\includegraphics[width=1.0\textwidth]{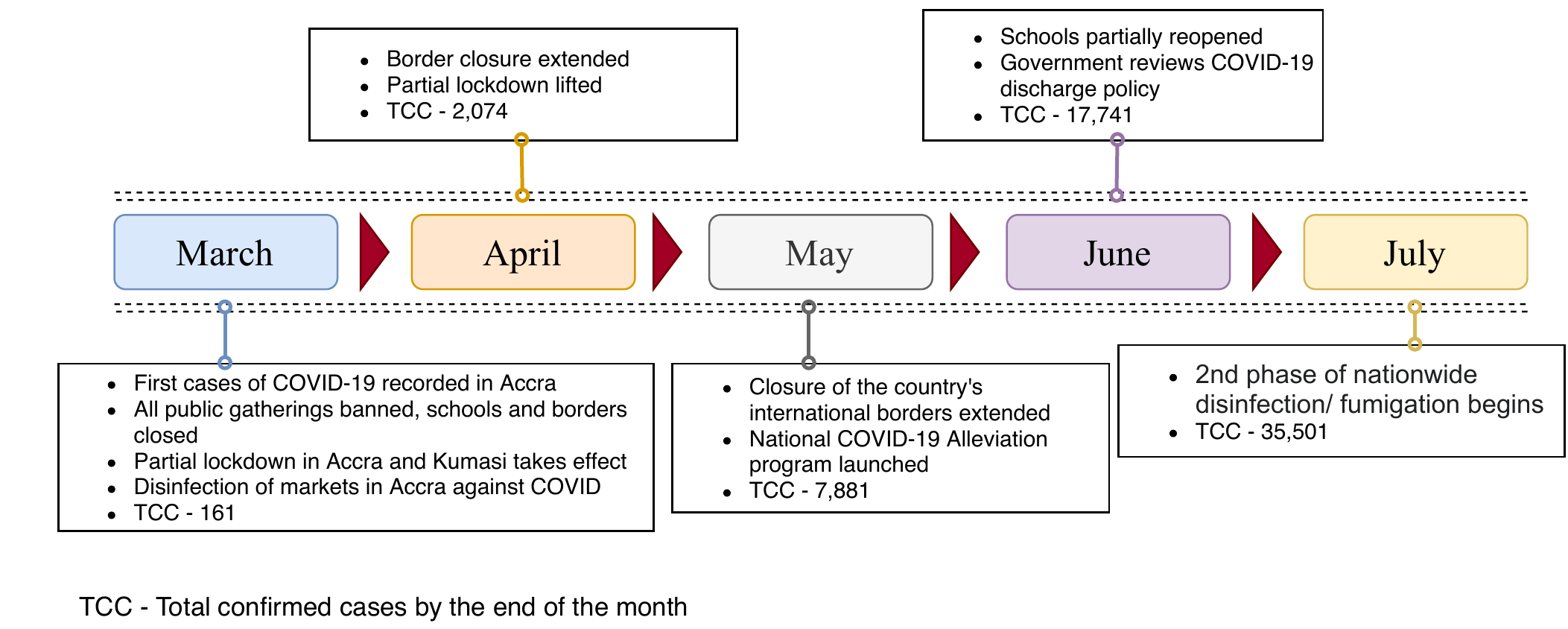}
}
\caption{Timeline of events in Ghana}
\label{fig:ga_0}
\end{figure}

Ghana had its first imported case of COVID-19 in Accra, the nation’s capital, and the regional capital of the Greater Accra Region, on 12-03- 2020 \cite{razakm},\cite{ASAMOAH}. The disease then rapidly spread across the other regions. As of 3-08-2020, 10:35 GMT+2, a total of 37,014 cases had been confirmed. The Greater Accra Region alone accounted for about 18,882, which is about 51\% of the total cases in Ghana\cite{ghswebsite}. Figure \ref{fig:ga_0} summarized the events from Ghana, etween March-July.

The coronavirus disease is caused by a pathogen that attacks the respiratory system of humans. Non-infected persons may contract the disease after physical contact with an infected person or contaminated surfaces and objects. The disease is characterized by fever, dry cough, and tiredness, which manifests within 14 days of infection. Other symptoms include diarrhea, sore throat, conjunctivitis, headache, and loss of taste and smell \cite{symtomswho}, \cite{Song1143}. 
To slow down and control the spread of the virus, The WHO defined some measures in the absence of an approved cure \cite{whoplan}. 

Many governments soon adopted and implemented these measures in their own countries. These measures include physical distancing between persons, wearing face masks, avoiding physical interactions like handshaking, and hugging, and lockdowns, etc\cite{Bonful2020.06.03.20120196}.
These measures have been communicated to individuals through various communication channels, including social media platforms like Twitter, Facebook, etc.

As a result, concerned individuals and public health officials are able to stay up to date with research findings, statistical and precautionary measures. Others use these platforms to share their own opinions and sentiments towards media reports or news\cite{meng2020}. A careful analysis of what individuals talk about will help satisfy digital epidemiological needs for efficient disease surveillance and case severity analysis. Healthcare providers or policymakers will then understand the state of the situation, thereby making it easy to reach out with available remedies\cite{9043580}.

In this work, we use data sourced from Twitter, which is the most used social network after Facebook and YouTube\cite{socialmediause} by a section of the Ghanaian population in the Greater Accra Region.

We contribute by providing a data-based assessment of the COVID-19 situation and the population’s response. This will aid healthcare providers and educators to assess how individuals at the target location feel, respond and understand matters concerning the COVID-19 pandemic. Based on this information, healthcare needs and decisions that are more suited to the population can be made and implemented.


\section{Proposal}

This section describes the proposal to analyze the situation in the country of Ghana in detail. The next step are part of the proposal:

\begin{itemize}
    \item Select keywords related to covid19 pandemic
    \item Set the parameters to collect related data
    \item Pre-processing text
    \item Visualization to support Analysis
\end{itemize}

\subsection{Select relevant terms}

The scope of the analysis is the Greater Accra region of Ghana. This choice was informed by the concentration of the population and access to the internet. The search terms used to retrieve information from the social network are:
\begin{itemize}
\item 'covid', 'coronavirus', 'sars'
\end{itemize}

\subsection{Build the Query and Collect Data}
\label{subsection:collection}

\begin{figure}[hbpt]
\centerline{
\includegraphics[width=0.4\textwidth]{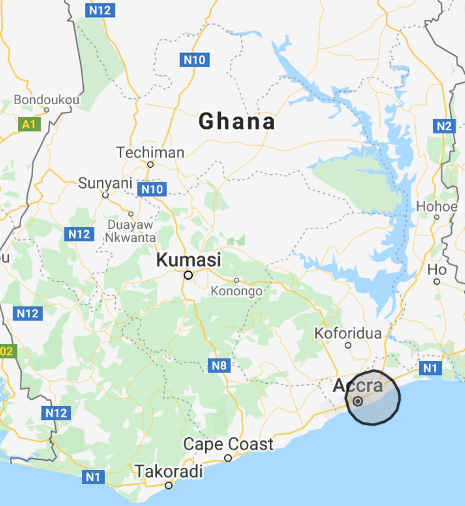}
\includegraphics[width=0.4\textwidth]{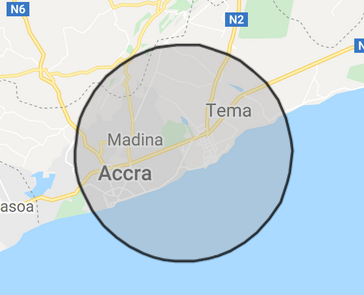}
}
\caption{Geolocalization of the Greater Accra Region, Ghana}
\label{fig:accra}
\end{figure}

Twitter is a social network with thousands of users sending micro-messages, up to 240 characters. Then, the collection process is through Twitter Search function using the parameters that follow:
\begin{itemize}
    \item date: 01-03-2020 to 15-07-2020
    \item terms: the chosen words mentioned in previous subsection
    \item geolocalization: 5.646129,-0.065209, see Fig. \ref{fig:accra}
    \item language: English
    \item radius: 25.3347 km
\end{itemize}

\subsection{Preprocessing Data}
In this step, the following procedures are taken to remove unwanted words and characters while maintaining the relevant ones. We also create visualisations to support our analysis and visual understanding. 

\begin{itemize}
    \item Uppercase to lowercase
    \item Eliminate alphanumeric symbols
    \item Remove stopwords, i.e. articles, pronouns, etc.
    \item Remove custom stopwords, i.e. coronavirus, covid, sars
\end{itemize}

The last step is useful to avoid coronavirus keywords and discover other terms related to the topic. According to the objective, this set of words can be chosen.

\subsection{Visualization}

After cleaning the data, the text is ready to create graphics to help the understanding of the situation in Ghana during the last months.

\begin{itemize}
\item Bar plots to show the frequency of tweets per day, per month
\item Cloud of words to visualize the most frequent terms per month
\item Line plots to show the trend of data
\end{itemize}

\section{Results}

This section explains the results obtained and the respective analysis. First subsection \ref{subsection:dataset} introduces the description of the dataset. 
Next subsections \ref{subsection:frequency}, \ref{subsection:terms} presents the frequency of post daily,monthly and \ref{subsection:terms}, the most frequent terms per month. Finally, a crossing graphic is presented in \ref{subsection:relationship} to analyze the relationship between Twitter, Google Trends, and new reported cases per day.

\subsection{Dataset}
\label{subsection:dataset}

This dataset is the result of the collection explained in \ref{subsection:collection} previously.

\begin{itemize}
    \item Total size of dataset: 9475 tweets
    \item Fields: date(YYYY-MM-DD), text(alphanumeric)
    \item Range Date: 01-03-2020 to 15-07-2020
    \item Language: English
\end{itemize}

\subsection{How is the frequency of post in the Greater Accra Region?}
\label{subsection:frequency}

\begin{figure}[hbpt]
\centerline{
\includegraphics[width=0.5\textwidth]{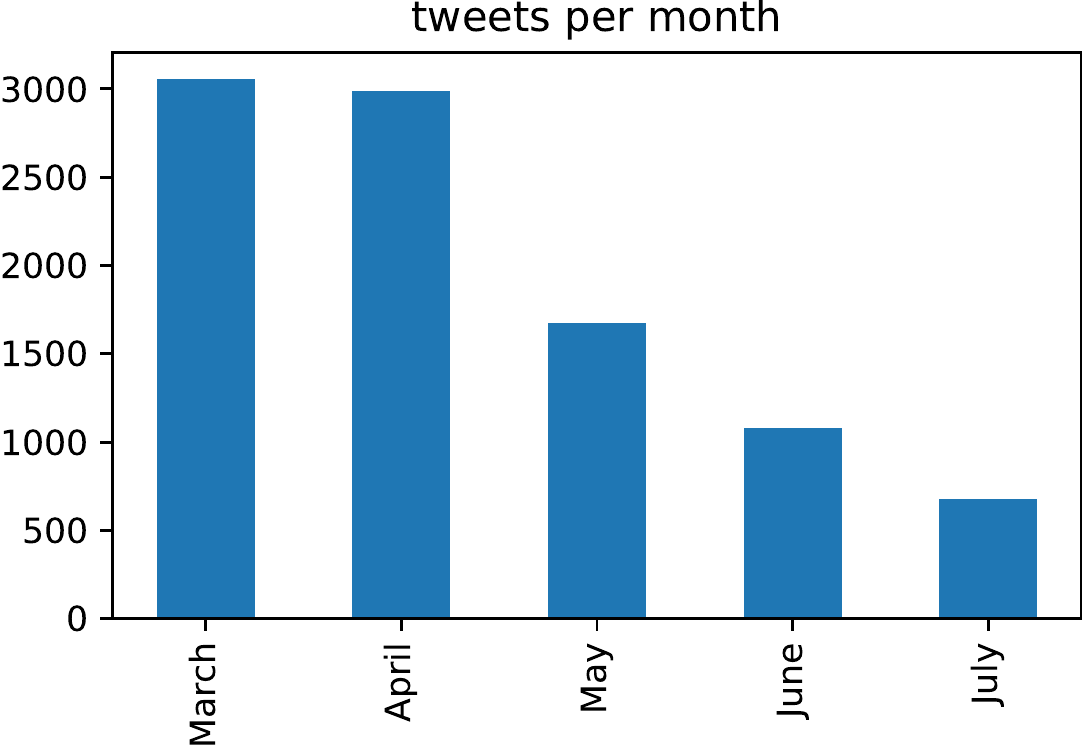}
}
\caption{Number of Tweets per month}
\label{fig:ga_1}
\end{figure}

First, it is important to know how many tweets/posts related to the topic of the study because this can express the interest of the users about the topic around covid-19 pandemic. Figure \ref{fig:ga_1}, shows March and April had the highest number of posts and from April there is less tweets month after month. 

\begin{figure}[hbpt]
\centerline{
\includegraphics[width=0.7\textwidth]{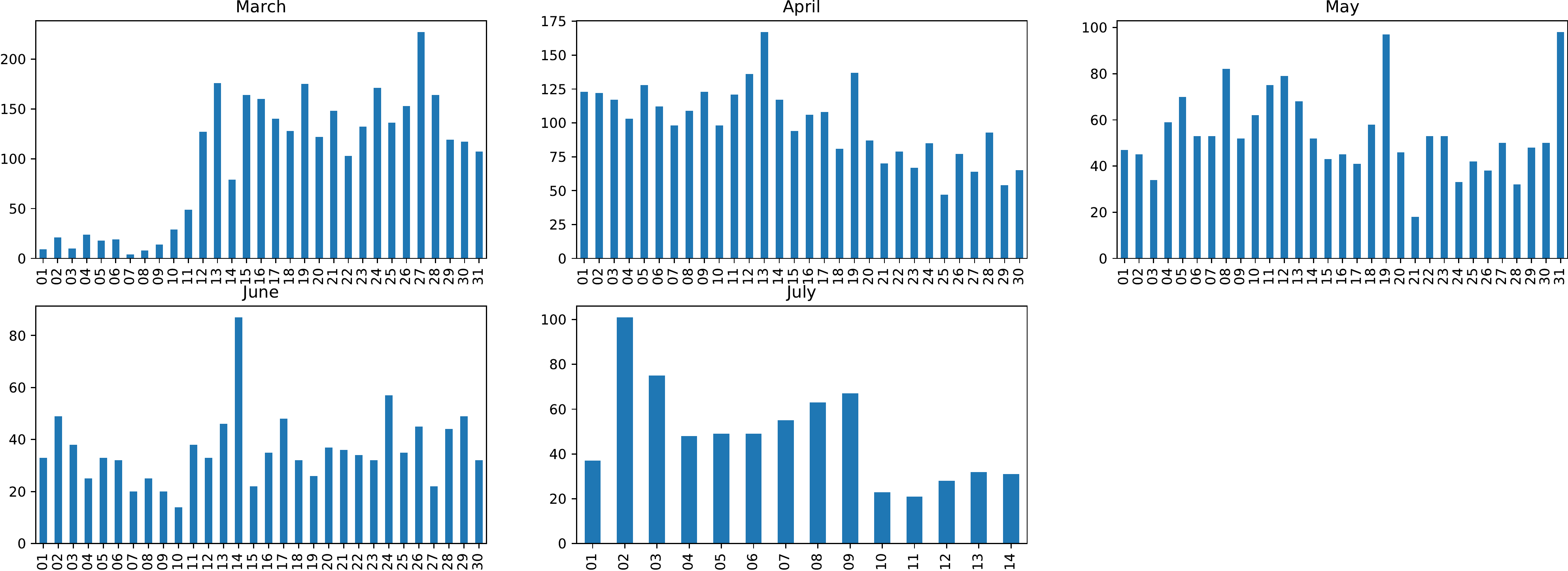}
}
\caption{Number of Tweets per day}
\label{fig:ga_2}
\end{figure}

Splitting the previous graphic, it is possible to show the frequency daily. Then, figure \ref{fig:ga_2} can summarize this information. Observing this graphic, from 01 to 27 March there is a constant increasing number of tweets but after 27 March, it starts to decrease. There are some peaks on March(27), April(13,19), May(19,31), June(14,24) and July(02), these numbers could be related to some events in the region.

The highlighted dates can be the start of an elaborated study around these days, then the study can be extended.

\subsection{Relevant terms during the months?}
\label{subsection:terms}

\begin{figure}[hbpt]
\centerline{
\includegraphics[width=0.6\textwidth]{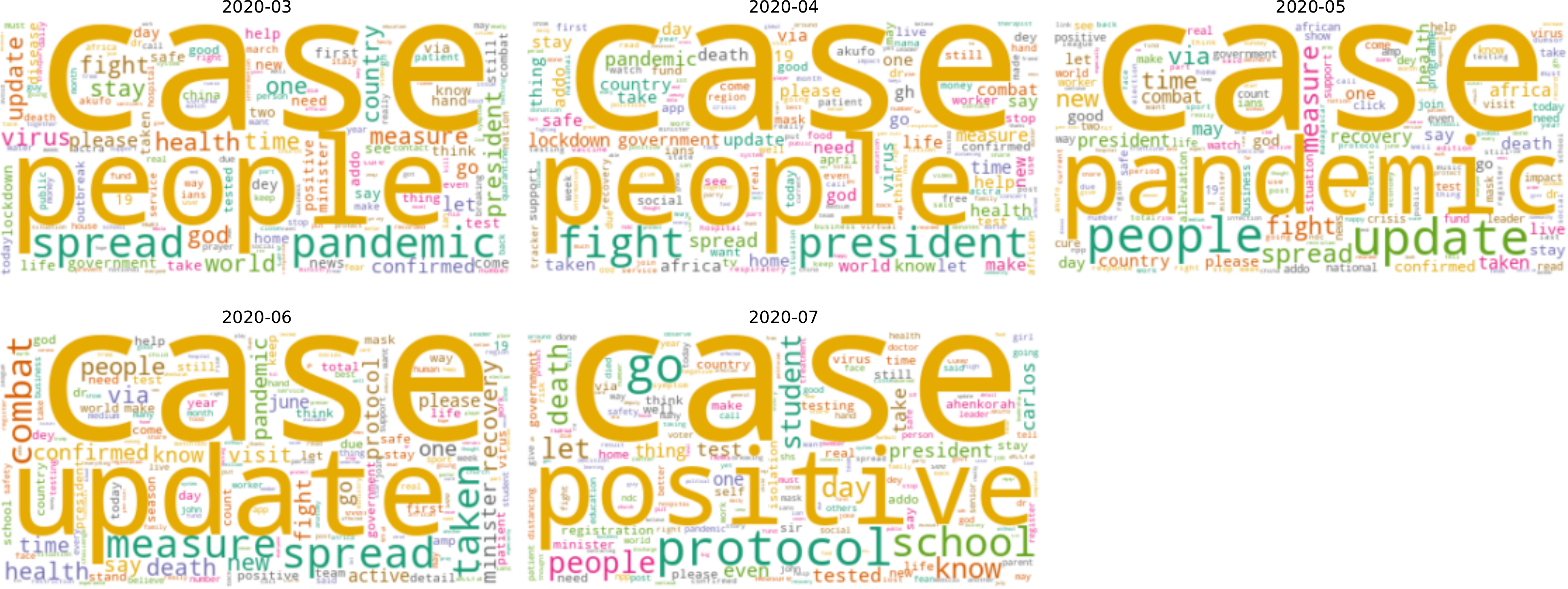}
}
\caption{Cloud of words - Raw}
\label{fig:ga_3}
\end{figure}

The terms: covid19, coronavirus, sars were extracted because the intention is to know the most frequent terms around these keywords per month. First, graphic \ref{fig:ga_3} shows case, people, pandemic, update, positive. Then, this keywords are more related to Public Health reports.

\begin{figure}[hbpt]
\centerline{
\includegraphics[width=0.8\textwidth]{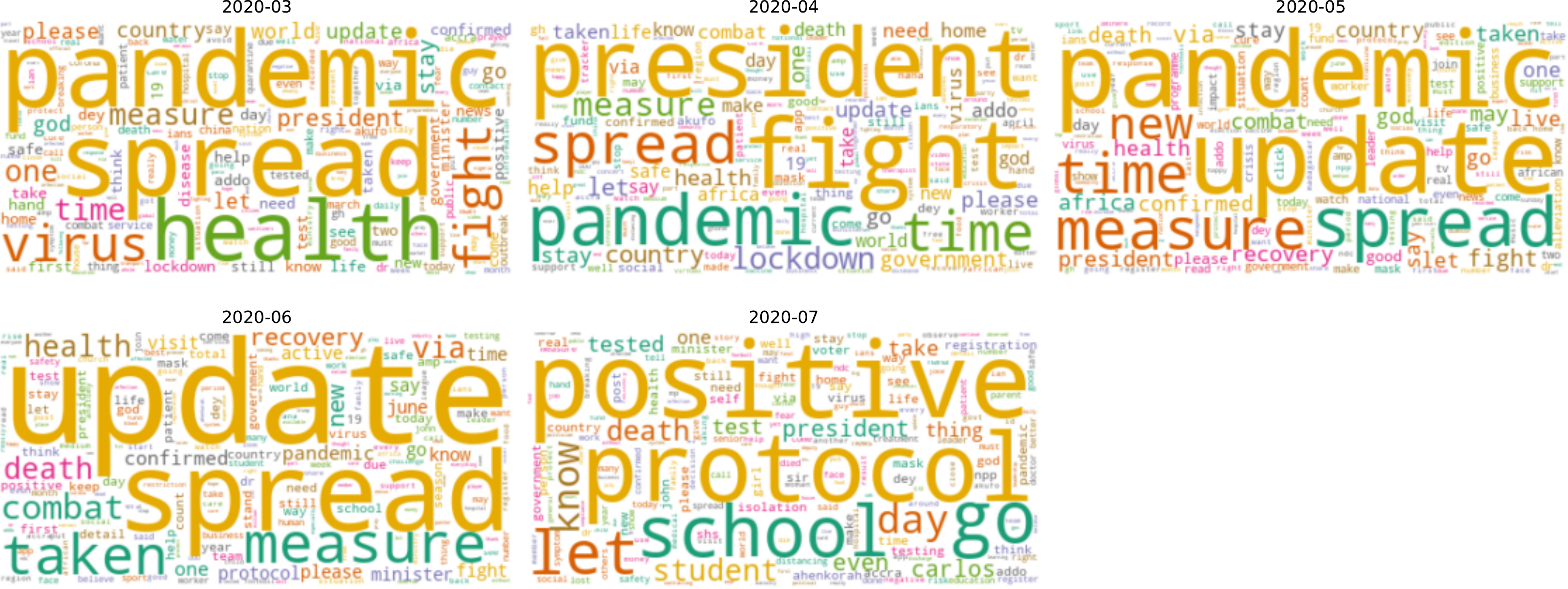}
}
\caption{Number of Tweets per day}
\label{fig:ga_4}
\end{figure}

Later, case and people terms are unconsidered and the result is graphic \ref{fig:ga_4}. Now, the most frequent are: pandemic, spread, health, president, update, protocol, school. Once more time, the posts are related to Public Health policies/actions of Ghana government.

Next figure \ref{fig:ga_5} presents the most frequent N-grams with N=2,3 to perform a better analysis about the previous terms. By consequence, it possible to confirm most of the tweets are related to actions of government, i.e. president akufo addo, stay home, taken combat spread, update measure taken.

\begin{figure}[hbpt]
\centerline{
\includegraphics[width=0.5\textwidth]{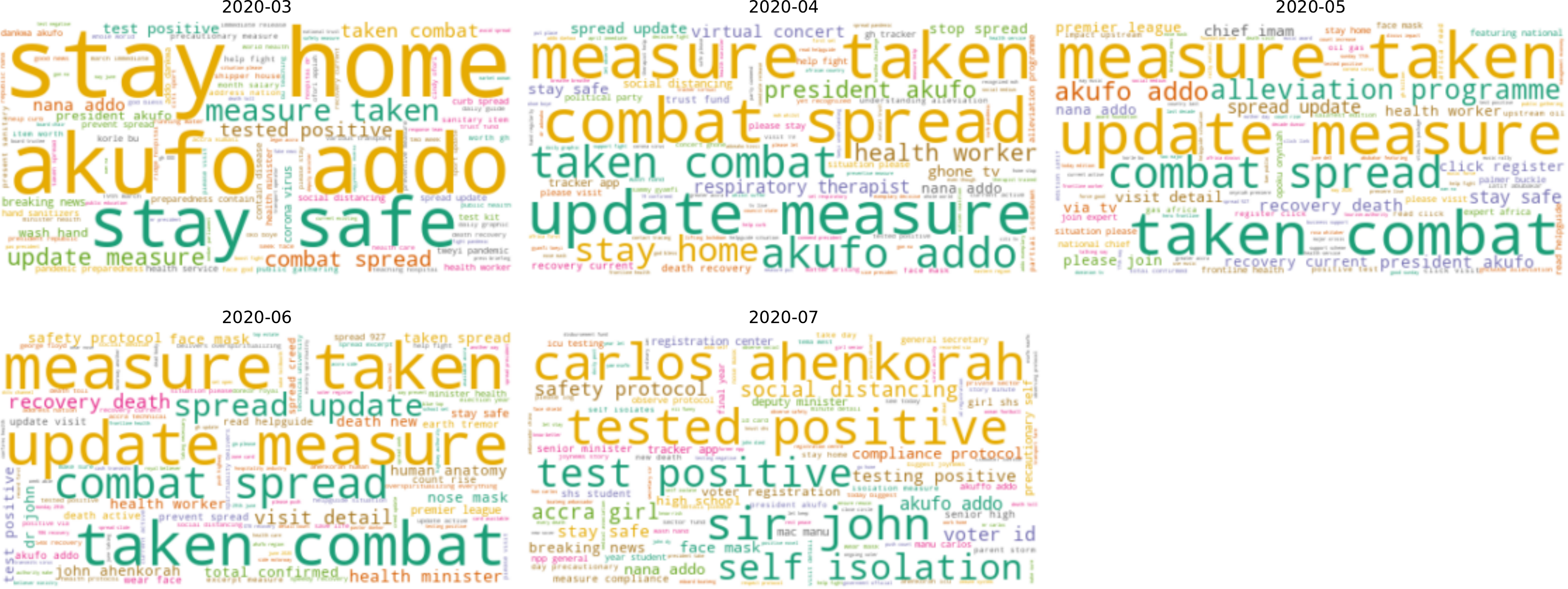}
\includegraphics[width=0.5\textwidth]{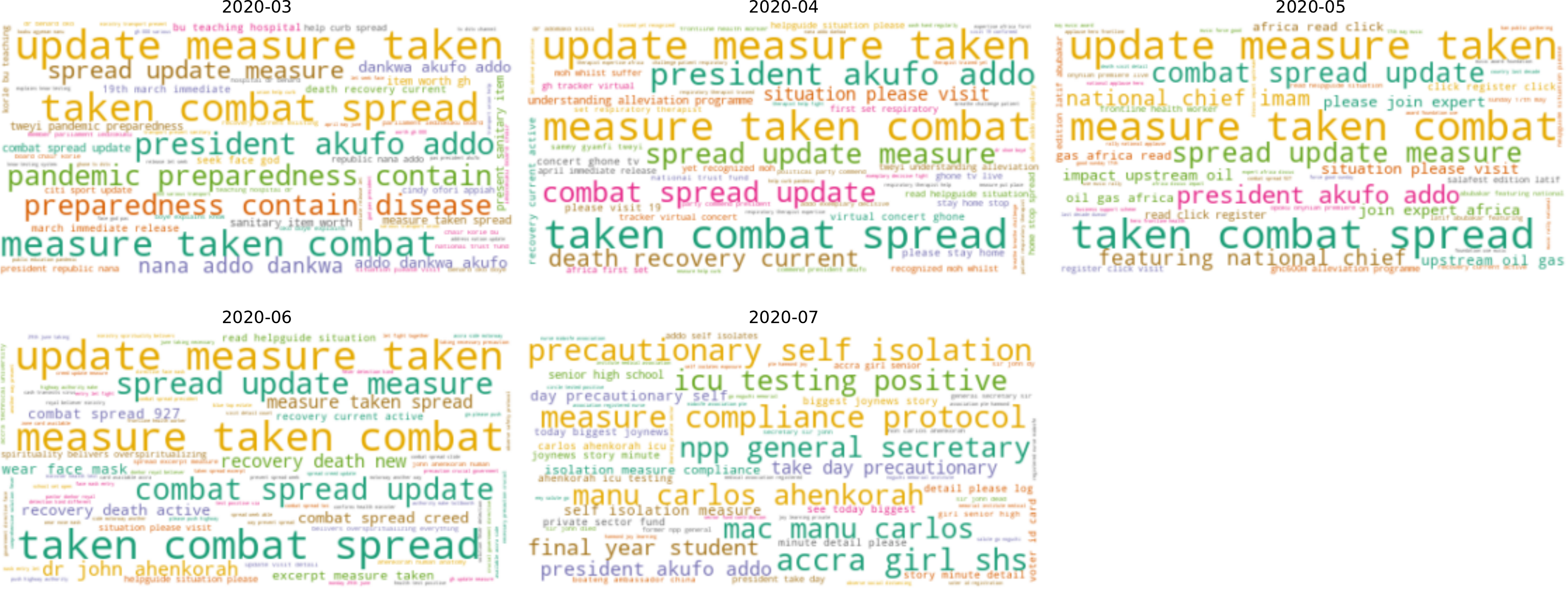}
}
\caption{Number of Tweets per day}
\label{fig:ga_5}
\end{figure}

\subsection{Understanding sentiments of the people}
The choice of words of users on social networks indicates their attitude and feeling about the situation they post about. In this section, we focus on words that conveys sentiments of the user from their composition. We categorized the pre-processed tokens into three namely; positive, negative and neutral based on their polarity scores.

\begin{figure}[hbpt]
\centerline{
\includegraphics[width=1.4\textwidth]{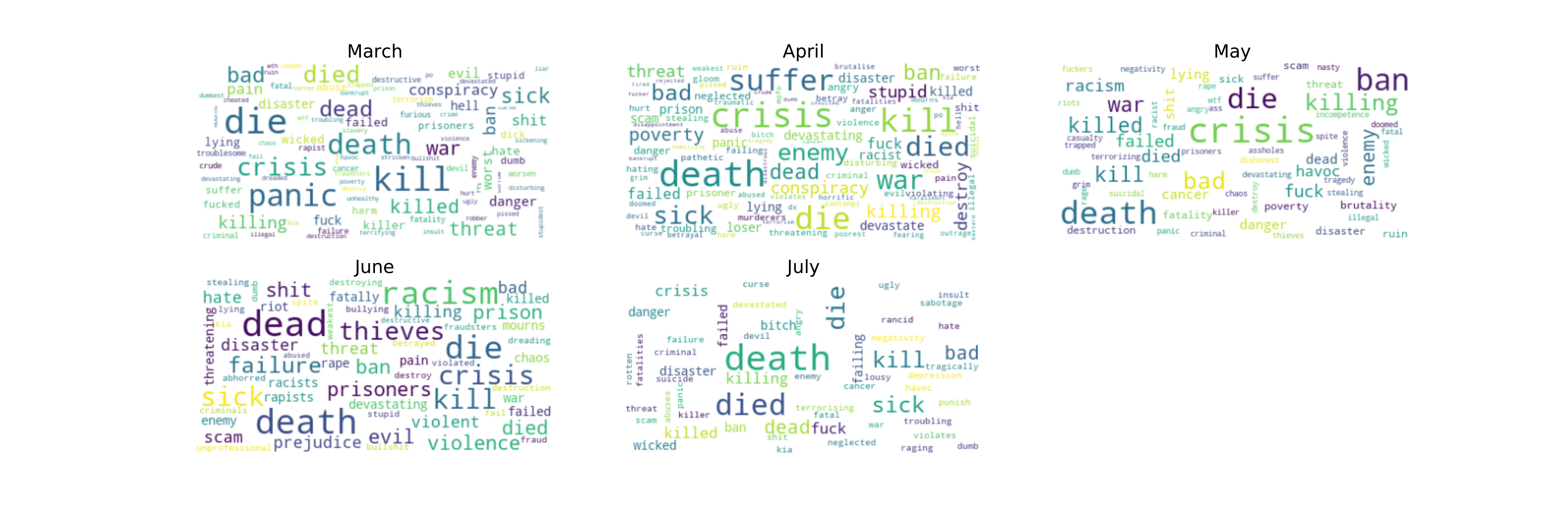}
}
\caption{Negative words of sentiments}
\label{fig:neg_sentiments}
\end{figure}

In figure \ref{fig:neg_sentiments}, frequently occurring words across months are ``death", ``die", ``kill". These words suggests that people were worried about the alarming death rate of the pandemic and the cause of the panic among them. Again, ``racism" appears in May and more frequently in June pointing out another cause of distress that likely surrounds the issues and behaviours against racial minorities.
\label{subsection:sentiments}

\subsection{What is the relationship between Twitter Posts, Daily Reported Cases and Google Trends?}
\label{subsection:relationship}

\begin{figure}[hbpt]
\centerline{
\includegraphics[width=0.9\textwidth]{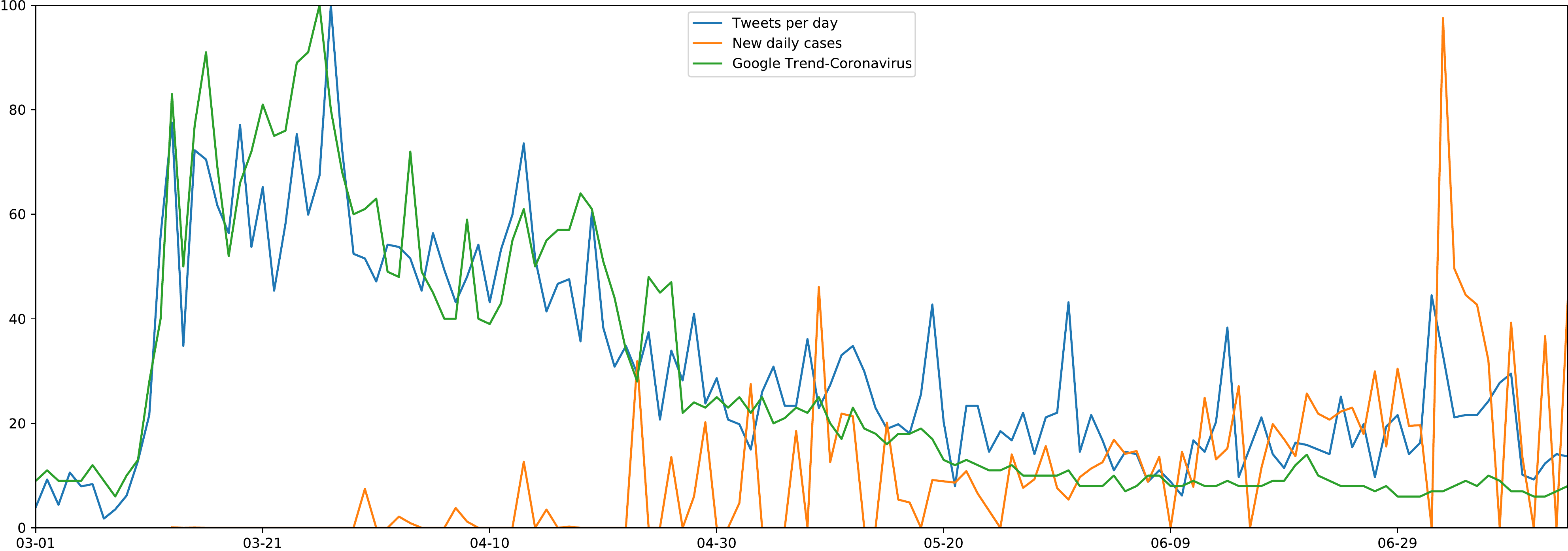}
}
\caption{Graphic Daily reported cases, Google Search, Twitter Posts}
\label{fig:ga_6}
\end{figure}

A valid question arises, how related can be posts of users with the actual pandemic? Therefore, graphic \ref{fig:ga_6} presents the daily reported new cases, interest of users about Google search and Twitter posts.

The scale of Twitter posts was adapted to have values between 0-100 like Google trends. The data to elaborate graphic related to daily cases was downloaded from HDX website \footnote{https://data.humdata.org/dataset/bc3589a6-04bc-4681-b531-7910ec800b4f}.

There is visual correlation between searches about coronavirus topic and messasges posted daily. The interest of user has decreased during the time, it can be the result of many social, mental effects produced by lockdown or constant fear facing the pandemic. An opposite results with a increasing number of cases in the country, people can be tired of the context and started to adapt and reactivate their lives. This analysis is open for future work.

\section{Conclusions}

First, Twitter can be a useful source of data. Process to analyze text includes the cleaning of the text. Second, there was an increasing interest on the pandemic as topic in Ghana user at the beginning, during the first reported cases. But, during the time, the interest about the pandemic were decreasing. The evidence for this affirmation is found in Google searches, Twitter posts daily, in spite of the number of new cases still is growing up.

\bibliographystyle{splncs04}
\bibliography{biblio.bib}

\end{document}